\documentclass[twocolumn,showpacs,prb,amsmath,amssymb,floatfix]{revtex4}
\usepackage{dcolumn}
\usepackage{amsmath,amssymb}
\usepackage{bm}
\usepackage{epsfig}
\usepackage{psfrag}

\newcommand{\bd}{\bm}

\begin{document}

\title{Thermal fluctuations of free standing graphene}
\author{F. L. Braghin} 
\author{N. Hasselmann}
\affiliation{International Institute of Physics, Univ. Fed. do Rio
Grande do Norte, 59072-970 Natal/RN, Brazil 
}
\date{26 March 2010}
\begin{abstract}
We use non-perturbative renormalization group 
techniques to calculate the 
momentum dependence of thermal fluctuations
of graphene,
based on a self-consistent calculation of the momentum
dependent elastic constants of a tethered membrane. 
We find a sharp crossover from
the perturbative to the anomalous regime,
in excellent agreement with
Monte Carlo results for 
the the out-of-plane fluctuations of 
graphene,
and give an accurate value for the crossover scale.
Our work strongly supports the notion that graphene is well
described as a tethered membrane. Ripples emerge naturally
from our analysis.

\end{abstract}
\pacs{05.10.Cc,63.22.Rc,68.60.Dv,68.65.Pq}
\maketitle
\section{Introduction}
Free standing graphene, the only presently known mono-atomic two-dimensional
crystal \cite{Castroneto09}, should be an ideal
tethered membrane, i.e. a membrane made
up of constituents with a fixed connectivity that give
rise to a finite shear modulus of the membrane. 
Tethered membranes
are known to have highly unusual properties, such as the
absence of any finite elastic constants in the thermodynamic
limit, a negative Poisson ratio, and fluctuations
characterized by a large anomalous dimension $\eta$ in the infra-red (IR) 
limit
\cite{membranebook}.
Experiments have not yet probed the large
wavelength regime
to test these
predictions, 
however a negative Poisson ratio and anomalous
fluctuations have been seen in Monte Carlo (MC) simulations
of graphene based on an  realistic effective many-body interaction
of C-atoms \cite{Los09}. One of the most surprising outcome of
experimental investigations of free standing graphene
were the observation of ripples in graphene
sheets with a characteristic scale $50-100\, {\text \AA}$ \cite{Meyer07}.
While it was often argued that these ripples 
are  not compatible with the standard continuum elastic theory
of tethered membranes \cite{Fasolino07,Gazit09b}, we demonstrate below
that they emerge naturally from it.
Since ripples are a finite scale
phenomenon, this requires to go beyond the asymptotic
regime which was investigated in previous theoretical
investigations.
Here, we present the most thorough RG treatment of tethered
membranes yet, 
a non-perturbative renormalization
group (NPRG) analysis of tethered
membranes which is based on the expansion of the effective
action in terms of elastic coupling {\em functions} which
for the first time allows to extract the full momentum
dependence of thermal fluctuations.  
Excellent agreement
with MC simulations of free standing graphene is found. 
Ripples emerge
as the real space analog of the Ginzburg scale, which
is the crossover scale which seperates the anomalous
regime from the perturbative one. We further calculate
the anomalous dimension both in the flat phase and
at the crumpling transition which is found to be of
second order.

In contrast to fluid membranes
which are always crumpled (the average of the normal of the surface vanishes)
the finite shear modulus stabilizes
a flat phase with long range order of the normals \cite{membranebook,Nelson87}
and 
the normal-normal correlation function $G_N$
decays asymptotically for small momenta $q$ as $G_N(q)~\simeq
q^{-2+\eta}$, see Refs.~[\onlinecite{membranebook,Paczuski88,Aronovitz89}].
The flat phase is stable for $\eta>0$ and in fact
all calculations yield a large anomalous dimension,
varying between $\eta=2/d=2/3$ from a large
$d$ expansion for $D$-dimensional membranes embedded in $d$-dimensional 
space \cite{Aronovitz89}, $\eta\approx 0.821$ from the
self-consistent screening approximation \cite{Doussal92,Gazit09} and
$\eta\approx 0.85$ where the last result was obtained both from 
NPRG \cite{Kownacki09} as well as from MC
simulations of graphene \cite{Los09}.
The 
analysis from Ref.~[\onlinecite{Kownacki09}]  
was able to reproduce all previously known results 
for general $d$,$D$
obtained via perturbative
RG methods within a unified framework. 
While the analysis in Ref.~[\onlinecite{Kownacki09}] is restricted to the 
leading order
of a derivative expansion of the action,
here we will significantly extend their
analysis in a way which allows to investigate thermal fluctuations at all momenta up
to the ultraviolet (UV) cutoff (we keep here only to the physical most 
relevant situation $d=D+1$ but our results
are easily extended to the general case).
Our analysis also yields
the correlation functions for the in-plane modes, but 
we confine the discussion
to $G_{hh}$ which for small $q$ is related to the normal-normal correlation function via $G_N\simeq q^2 G_{hh}$.\cite{membranebook}
However,
since $G_{hh}$ is 
more readily accessible than $G_N$ in the NPRG approach,
 we will base our analysis on $G_{hh}$.

\section{The model}
We start from a Landau-Ginzburg type
ansatz \cite{membranebook,Paczuski88,Aronovitz89,Kownacki09} for the
energy functional of a tethered membrane ${\cal H}={\cal H}^{\rm
b}+{\cal H}^{\rm st}$ which consists of a bending part
\begin{subequations}
\begin{equation}
{\cal H}^{\rm b}=\frac{\tilde\kappa}{2}\int d^D{x} \left(\partial_a\partial_a {\bd R}\right)^2
\label{eq:bend}
\end{equation}
and a stretching part
\begin{eqnarray}
{\cal H}^{\rm st}&=&\int d^D x \big[\frac{{\tilde r}_0}{2}(\partial_a {\bd R})^2+
\frac{\tilde{\mu}}{4}
(\partial_a {\bd R}\cdot \partial_b {\bd R})^2 \nonumber \\
&&+\frac{\tilde{\lambda}}{8}
(\partial_a {\bd R}\cdot \partial_a {\bd R})^2 \big] \, ,
\label{eq:stretch}
\end{eqnarray}
\end{subequations}
where ${\bd R}$ is a $D+1$ dimensional vector parametrizing the $D$
dimensional membrane which is embedded in a $D+1$ dimensional space.
The presence of an UV cutoff $\Lambda_0$ is implicitly
assumed. The inverse temperature $\beta=1/k_B T$ is absorbed in the definition of the
effective parameters, i.e. $\tilde{\kappa}=\beta \kappa$,
$\tilde{\mu}=\beta \mu$, $\tilde{\lambda}=\beta \lambda$ and
${\tilde r}_0=\beta r_0$.
If one writes the stretching part of the membrane in
terms of derivatives, $\bd{m}_a=\partial_a \bd{R}(\bd{x})$,
$a=1\dots D$,
the analogy to a Ginzburg-Landau expansion becomes apparent and one
would expect a phase transition near ${\tilde r}_0\approx 0$ from a
symmetric, crumpled phase with $\big< {\bd m}_a \big>=0$
which exists for positive ${\tilde r}_0$ to a symmetry broken flat
phase, characterized by the order parameter $\big< {\bd m}_a \big>
={\bd m}_{a,0} =J \bd{e}_a \neq 0$ where $J$ is the magnitude of the
order parameter and $\bd{e}_a$, $a=1 \dots D$,  are unit vectors
which span the membrane,  $\big< \bd{R} \big>=J x_a
\bd{e}_a$.

 Here we shall be interested in the flat phase and therefore  rewrite Eq.~(\ref{eq:stretch})
by introducing the flat metric tensor
$g_{ab}^0=\bd{m}_{a,0}\cdot\bd{m}_{b,0}=J^2 \delta_{ab}$.  Defining  $g_{ab}=\bd{m}_a\cdot\bd{m}_b$
and $U_{ab}=(g_{ab}-g_{ab}^0)/2$, we find, up to constant
\cite{Aronovitz89},
\begin{eqnarray}
{\cal H}^{\rm st}&=&\int d^Dx \Big[\tilde{\mu}\, 
U_{ab}^2 +\frac{\tilde{\lambda}}{2} \,  
U_{aa}^2
\Big]
\label{eq:flatstretch}
\end{eqnarray}
where we used the mean field result $J_{\Lambda_0}=[-\tilde{r}_0/({\tilde \mu}
+D{\tilde \lambda}/2)]^{1/2}$ for the order parameter to cancel a
term linear in $U_{aa}$. It is convenient to separate in-plane and
out-of-plane deformations of the membrane by introducing $\Delta{\bd
m}_a=\partial_a\bd{R}-\bd{m}_{a,0}$ with $\Delta{\bd m}_a=(\partial_a {\bd
u},\partial_a h)$ such that $h$ corresponds to out-of-plane
deformations and ${\bd u}=u^a {\bd e}_a$. In these variables we have
$U_{ab}=(1/2)(J \partial_a u^b+J\partial_b u^a+\partial_a h
\partial_b h +
\partial_a {\bd u} \cdot \partial_b {\bd u})$. Note that if one
ignores terms of third and fourth order in
${\bd u}$ and keeps derivatives of ${\bd u}$ only up
to second order, one finds, after a rescaling 
\cite{Paczuski88,Aronovitz89,Kownacki09} 
$(h,{\bd u}) \to J (h,{\bd u})$,
the 
minimal model
for a flat
membrane \cite{Aronovitz88},
$
{\cal H}\approx \frac{1}{2}\int d^D x \big[ \bar{\kappa} (\partial_a \partial_a h)^2 +
\bar{\mu} u_{ab}^2 +\frac{\bar{\lambda}}{2} u_{aa}^2 \big]
$
with $u_{ab}=\partial_a u^b+\partial_b u^a + \partial_a h \partial_b h$ and ${\bar \mu}=\tilde{\mu} J^4$,
${\bar \lambda}=\tilde{\lambda} J^4$ and ${\bar \kappa}={\tilde \kappa}J^2$. The minimal model does
however not possess the full symmetry of the Ginzburg-Landau model defined by
Eqs.~(\ref{eq:bend}, \ref{eq:stretch}) and cannot describe the crumpling
transition. Furthermore, neglecting the fourth order derivative
terms of ${\bd u}$ prevents an accurate description of the membrane
fluctuations at finite momenta.  
We therefore do not use the minimal model here.

\section{Non-perturbative RG approach}
The NPRG is based on the exact flow equation  \cite{Wetterich93} 
for the cutoff dependent
effective action $\Gamma_\Lambda$ which for $\Lambda=\Lambda_0$ coincides
with the bare action $\cal H$,
\begin{equation}
\frac{\partial \Gamma_\Lambda}{\partial \Lambda}=
\frac{1}{2} {\rm Tr} \left[ \left( \frac{\partial^2\Gamma_\Lambda}{\partial \phi \partial \phi^\prime}
+R_\Lambda \right)^{-1} \frac{\partial R_\Lambda}{\partial \Lambda} \right] ,
\end{equation}
where $\Lambda$ is the running IR cutoff and $\phi$, $\phi^\prime$ are any of the fields $u^a$ or $h$.
The trace stands for an integral over momentum and a sum over internal indices.
The regulator function $R_\Lambda$ removes IR divergences arising from modes with $k<\Lambda$ and
will be specified below.
The NPRG analysis of Ref.~[\onlinecite{Kownacki09}] was restricted to the flow of the parameters which appear already in $\Gamma_{\Lambda_0}$.
While this is 
sufficient to discuss the asymptotic regime at vanishingly small momenta,
the RG transformation will in general lead to a momentum dependence of
$\tilde{\kappa}$, $\tilde{\mu}$, and
$\tilde{\lambda}$ which must be accounted for in a proper analysis
of thermal fluctuations at finite momenta. For
$\Lambda<\Lambda_0$ we shall therefore make a non-local ansatz 
of the form $\Gamma_\Lambda=\Gamma_\Lambda^{\rm b}
+ \Gamma_\Lambda^{\rm st}$ with
\begin{subequations}
\begin{equation}
\Gamma_\Lambda^{\rm b}= \frac{1}{2} \int d^Dx \ d^Dx^\prime {\tilde \kappa}_\Lambda({\bd x}-{\bd x}^\prime)
\partial_a^2 {\bd R}({\bd x}) \partial_b^2 {\bd R}({\bd x}^\prime)
\label{eq:nlbend}
\end{equation}
and
\begin{eqnarray}
\Gamma_\Lambda^{\rm st}&=&\int d^Dx \ d^Dx^\prime \Big[ {\tilde \mu}_\Lambda({\bd x}-{\bd x}^\prime)
U_{ab}({\bd x})U_{ab}({\bd x}^\prime) \nonumber
\\ && +\frac{1}{2}{\tilde \lambda}_\Lambda({\bd x}-{\bd x}^\prime) 
U_{aa}({\bd x})U_{bb}({\bd x}^\prime) \Big] .
\label{eq:nlstretch}
\end{eqnarray}
\end{subequations}
This rather natural generalization of the effective action allows to account for non-local correlations
but at the same time ensures that, as long as only the coupling {\em functions} ${ \tilde \kappa}_\Lambda$,
${\tilde \mu}_\Lambda$ and ${\tilde \lambda}_\Lambda$ and the parameter $J_\Lambda$ are renormalized,
the effective action retains at all $\Lambda$ the full symmetry of the original model and thus all
Ward identities will be obeyed. Apart from the approximation that we only take into account the irreducible
correlations explicitly defined through Eqs.~(\ref{eq:nlbend},\ref{eq:nlstretch}), which uniquely fixes
the RG flow equations, no further approximations will be made.

\begin{figure}[h]
\includegraphics[width=7.5cm]{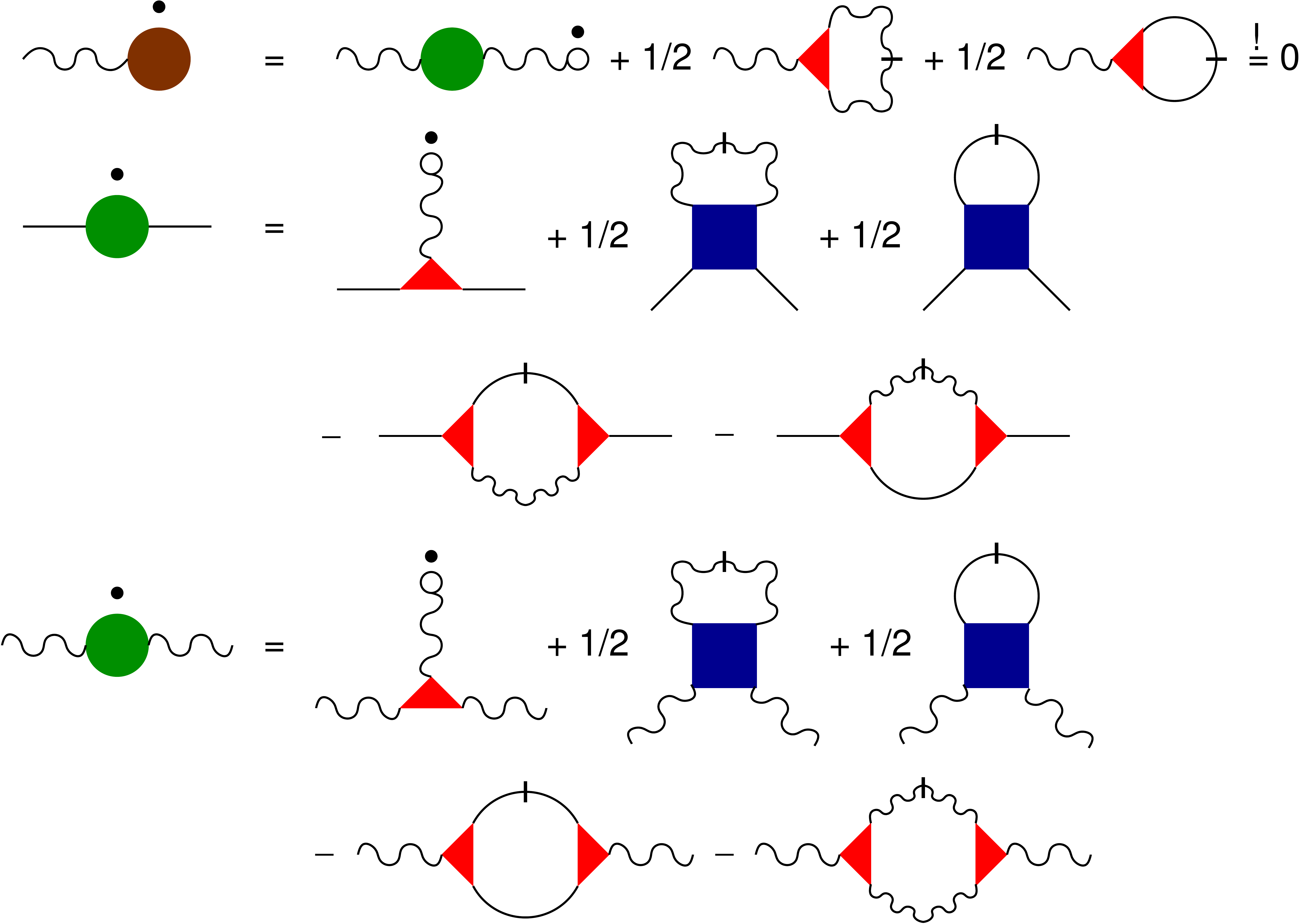}
\caption{(Color online) 
Flow diagrams for the one point vertex (1st line) and the
self-energies $\Sigma_{hh}$ (2nd line) and $\Sigma_{ab}$. Forcing
the one-point vertex to vanish for all $\Lambda$
yields the flow of the order parameter $J_\Lambda$.\cite{Schuetz06} Wiggly
lines correspond to $G_{ab}$ propagators and solid lines to
$G_{hh}$. Lines with a dash correspond to single-scale
propagators defined via ${\dot G}_{ab}=-G_{ab}^2\partial_\Lambda
R_{\Lambda}$ and ${\dot G}_{hh}= -G_{hh}^2\partial_\Lambda
R_{\Lambda}$. Small open circles denote the order parameter and the
solid dot a derivative with respect to $\Lambda$. } \label{fig:diagrams}
\end{figure}

To derive the NPRG equations we expand $\Gamma_\Lambda$ in the fields 
$\Delta{\bd m}_a
=(\partial_a {\bd u},\partial_a h)$.
The Dyson equation for the Greens function $G_{hh}$ of the $h$ field, defined via $\big< h_{\bd q} h_{-{\bd q}^\prime}\big>=
V \delta_{{\bd q},{\bd q^\prime}} G_{hh}(q)$ where $V$ is the $D$-dimensional volume, is
\begin{equation}
G_{hh}^{-1}(q)=G_{0,\Lambda}^{-1}(q)+\Sigma_{hh}(q)
\end{equation}
with (here and below we suppress in our notation the
$\Lambda$ dependence of the coupling parameters)
\begin{subequations}
\begin{eqnarray}
\Sigma_{hh}(q)&=&({\tilde \kappa}_q-{\tilde \kappa}_{\Lambda_0})q^4 , \\
G_{0,\Lambda}^{-1}&=&{\tilde \kappa}_{\Lambda_0} q^4+R_\Lambda (q),
\end{eqnarray}
\end{subequations}

\noindent where ${\tilde \kappa}_{\Lambda_0}$ denotes the bare and momentum independent value of the
initial coupling constant ${\tilde \kappa}$ defined at the UV cutoff $\Lambda_0$.
The Greens functions of the in-plane modes, defined via
$\big<u^a_{\bd k} u^b_{-{\bd k}^\prime}\big>=V \delta_{{\bd k},{\bd k}^\prime} G_{ab}({\bd k})$,
can be written in terms of transverse and longitudinal components,
\begin{equation}
G_{ab}({\bd k}) = G_\perp(k)\left(\delta_{ab}-k^a k^b/k^2 \right)+
G_\parallel(k) k^a k^b/k^2
\end{equation}

\noindent with
$G_{\alpha}^{-1}= G_{0,\Lambda}^{-1}+\Sigma_\alpha$ for $\alpha=\perp,\parallel$, and

\begin{subequations}
\begin{eqnarray}
\Sigma_\perp(k)&=&J^2 {\tilde \mu}_k k^2+({\tilde \kappa}_k-{\tilde \kappa}_{\Lambda_0})k^4 , \\
\Sigma_\parallel(k)&=&J^2 (2{\tilde \mu}_k+{\tilde \lambda}_k) k^2+
({\tilde \kappa}_k-{\tilde \kappa}_{\Lambda_0})k^4 .
\end{eqnarray}
\end{subequations}

\noindent
To determine the flow of $J_\Lambda$ and the self-energies, we further need the three and
four point vertices. 
In a symmetrized form they read

\begin{subequations}
  \begin{widetext}
    \begin{eqnarray}
    \label{eq:g3}
      \Gamma^{(3)}_{abc}(\bd{k}_1,\bd{k}_2,\bd{k}_3)&=&    -iJ \Big\{
      \tilde{\lambda}_{k_3} (\bd{k}_1\cdot \bd{k}_2) k_3^c \delta_{ab}
    + \tilde{\lambda}_{k_2} (\bd{k}_1\cdot \bd{k}_3) k_2^b \delta_{ac}
      +\tilde{\lambda}_{k_1} (\bd{k}_2\cdot \bd{k}_3) k_1^a \delta_{bc}
      +
        (\bd{k}_1\cdot \bd{k}_2) ( \tilde{\mu}_{k_1} k_3^a \delta_{bc}+\tilde{\mu}_{k_2}k_3^b \delta_{ac})
    \nonumber
      \\ &&
      +(\bd{k}_1\cdot \bd{k}_3) (\tilde{\mu}_{k_1} k_2^a \delta_{bc}+\tilde{\mu}_{k_3}k_2^c \delta_{ab})
      +(\bd{k}_2\cdot \bd{k}_3) ( \tilde{\mu}_{k_2}k_1^b \delta_{ac}+\tilde{\mu}_{k_3}k_1^c \delta_{ab})
    \Big\},
      \\
      \Gamma^{(3)}_{hha}(\bd{q}_1,\bd{q}_2;\bd{k})&=&-iJ\Big\{
      \tilde{\mu}_k \big[(\bd{q}_1\cdot \bd{k}) q_2^a +(\bd{q}_2\cdot \bd{k} ) q_1^a\big]
      + \tilde{\lambda}_k (\bd{q}_1\cdot \bd{q}_2 ) k^a \Big\} ,
      \\
      \Gamma^{(4)}_{abcd}(\bd{k}_1,\bd{k}_2,\bd{k}_3,\bd{k}_4)&=&
      \tilde{\mu}_{k_{12}} [(\bd{k}_1\cdot \bd{k}_3)( \bd{k}_2\cdot \bd{k}_4)
    +(\bd{k}_1\cdot \bd{k}_4 )(\bd{k}_2\cdot \bd{k}_3)] \delta_{ab}\delta_{cd}
      +\tilde{\mu}_{k_{13}}[(\bd{k}_1\cdot \bd{k}_2)( \bd{k}_3\cdot \bd{k}_4)
    +(\bd{k}_1\cdot \bd{k}_4 )(\bd{k}_2\cdot \bd{k}_3)] \delta_{ac}\delta_{bd} \nonumber
      \\ &&
      +\tilde{\mu}_{k_{14}}[(\bd{k}_1\cdot \bd{k}_2 )(\bd{k}_3\cdot \bd{k}_4)
    +(\bd{k}_1\cdot \bd{k}_3)( \bd{k}_2\cdot \bd{k}_4)] \delta_{ad}\delta_{bc}  \nonumber
      \\ &&
      + \tilde{\lambda}_{k_{12}} (\bd{k}_1\cdot \bd{k}_2 )(\bd{k}_3\cdot \bd{k}_4) \delta_{ab} \delta_{cd}
      + \tilde{\lambda}_{k_{13}} (\bd{k}_1\cdot \bd{k}_3 )(\bd{k}_2\cdot \bd{k}_4) \delta_{ac} \delta_{bd}
      + \tilde{\lambda}_{k_{14}} (\bd{k}_1\cdot \bd{k}_4 )(\bd{k}_2\cdot \bd{k}_3) \delta_{ad} \delta_{bc} ,
      \\
      \Gamma^{(4)}_{hhab}(\bd{q}_1,\bd{q}_2;\bd{k}_1,\bd{k}_2)&=&
      \delta_{ab}\Big\{ \tilde{\mu}_{q_{12}} \big[(\bd{q}_1 \cdot \bd{k}_1)(\bd{q}_2 \cdot \bd{k}_2) +
      (\bd{q}_1 \cdot \bd{k}_2)(\bd{q}_2 \cdot \bd{k}_1)\big] +\tilde{\lambda}_{q_{12}} (\bd{q}_1 \cdot \bd{q}_2)
      (\bd{k}_1 \cdot \bd{k}_2)
      \Big\} \, ,
      \\
      \Gamma^{(4)}_{hhhh}(\bd{q}_1,\bd{q}_2,\bd{q}_3,\bd{q}_4)&=&
      (\tilde{\mu}_{q_{12}}+ \tilde{\mu}_{q_{14}}+\tilde{\lambda}_{q_{13}})
    (\bd{q}_1\cdot \bd{q}_3) (\bd{q}_2\cdot \bd{q}_4)
      +
      (\tilde{\mu}_{q_{12}}+ \tilde{\mu}_{q_{13}}+\tilde{\lambda}_{q_{14}})
    (\bd{q}_1\cdot \bd{q}_4)( \bd{q}_2\cdot \bd{q}_3) \nonumber \\ &&
      +
     (\tilde{\mu}_{q_{13}}+ \tilde{\mu}_{q_{14}}+\tilde{\lambda}_{q_{12}})
    (\bd{q}_1\cdot \bd{q}_2)( \bd{q}_3\cdot \bd{q}_4) ,
    \label{eq:g4}
    \end{eqnarray}
  \end{widetext}
\end{subequations}

\noindent
where $k_{ij}=|{\bd k}_i+{\bd k}_j|$. The subscript $h$ and momenta
${\bd q}_i$ refer to $h$ fields while subscripts $a \dots d$ and
${\bd k}_i$ refer to ${\bd u}$ fields. The flow equations for the
order parameter $J$ and the self-energies $\Sigma_{hh}$,
$\Sigma_\perp$, and $\Sigma_{\parallel}$ are rather long 
and shown diagrammatically in
Fig.~\ref{fig:diagrams}. They yield coupled integro-differential
equations for the flow of the coupling functions which must be
solved self-consistently. Note that the equations are
closed, since all three- and four-point vertices in  
Eqs.~(\ref{eq:g3}-\ref{eq:g4})
are entirely determined by
coupling functions which can be extracted 
from the
self-energies. This is a result of the
form of $\Gamma_\Lambda$ in
Eqs.~(\ref{eq:nlbend},\ref{eq:nlstretch}) 
which relates the third and fourth order vertices 
to lower order ones which in a usual field
expansion would have to be imposed through Ward identities. 

We integrate the flow equations from the UV cutoff
$\Lambda=\Lambda_0$ to $\Lambda=0$ numerically, using for numerical
stability an analytic regulator, $R_\Lambda(q)={\tilde
\kappa}_{q=0}\, q^4/( \exp[ (q/\Lambda)^4] -1)$ (this one, as well as a
non-analytic regulator, were also used in \cite{Kownacki09}). If we
ignore the momentum dependence of ${\tilde \kappa}_q$, ${\tilde
\mu}_q$, and ${\tilde \lambda}_q$ our flow equations reduce to those
reported in Ref.~[\onlinecite{Kownacki09}].
In the flat phase we find for $D=2$ $\eta\approx 0.85$ which
agrees with the derivative expansion result \cite{Kownacki09}. 
For completeness, we note that our NPRG approach
yields for $D=2$ a second order crumpling transition (to within numerical
accuracy) with an anomalous dimension $\eta\approx 0.64(5)$,
slightly larger than the result $\eta\approx 0.627$ 
obtained with a sharp cutoff
and a derivative expansion in \cite{Kownacki09}, where a weak
dependence of the flow on the form of the regulator prevented
a firm conclusion on the order of the transition.
Our result, obtained with a smooth regulator, resolves 
this ambiguity in favor of a transition of 
second order. However, we cannot rule out that 
terms of higher order in the stress tensor
would change the nature of the transition.

\section{Comparison with MC data and the role of the Ginzburg scale}

\begin{figure}[h]
\includegraphics[width=5.5 cm,angle=-90]{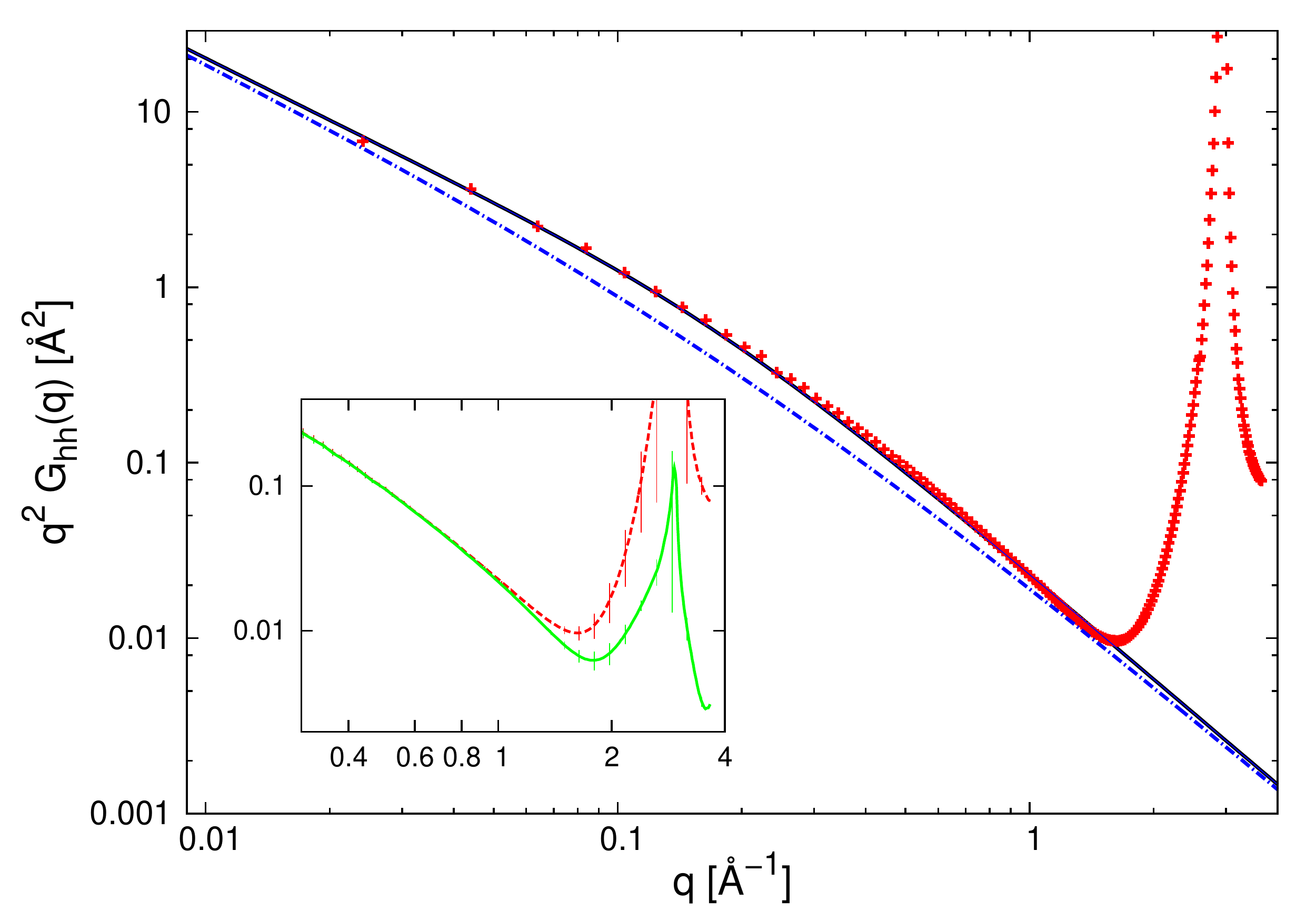}
\caption{(Color online)
Results for the out-of-plane fluctuations $q^2 G_{hh}(q)$ from
NPRG (solid, black), MC (dashed, red) and from a simple 
phenomenological approximation
discussed in the main text
(dashed-dotted, blue). The inset shows MC data for the out-of-plane
correlations $q^2 G_{hh}(q)$ (dashed, red) and the fluctuations of normals $G_N(q)$
(solid, green).
}
\label{fig:Ghh}
\end{figure}

The MC data for $G_{hh}$ were obtained from a system of $37888$
C-atoms with an accurate bond order potential LCBOPII and $T=300$~K,
see 
Refs.~\cite{Fasolino07,Los09,Zackarchenko09,Zackarchenko10}
for
details. To reduce statistical noise, the out-of-plane
distortions $h_i$ were obtained by evaluating 
for each $i$ the average ${\bar h}_i= ( 3 h_i
+\sum_{\left< i,j\right>}h_j)/6$ where the sum runs over the three
neighbors of atom $i$.\cite{Zackarchenko10} 
As expected, for small $q$ one finds the
relation $q^2 G_{hh}\simeq G_N$ 
between the correlation function of
the normals and the height fluctuation 
which for graphene 
is extremely accurate even up to $q\approx 1 \,
{\text \AA}^{-1}$, see the inset in Fig.~\ref{fig:Ghh}. Since the very
small $q$ data for $G_{hh}$ is more noisy than that of $G_N$ we used
for the last three data points ($q<0.07\, {\text \AA}^{-1}$) the data of
$G_N$ to calculate $G_{hh}$, the 
result 
is shown in
Fig.~\ref{fig:Ghh}. The strong peak near $q_B=4 \pi/{3 a}\simeq 2.94 \,
{\text \AA}^{-1}$, where $a$ is the equilibrium lattice parameter,
corresponds to the first Bragg peak. It defines the upper limit
beyond which continuum theory is inapplicable and it serves as a
natural UV cutoff for the NPRG calculation, $\Lambda_0=q_B$. For
smaller $q$ the data shows the scaling $G_{hh}\propto
q^{-4}$ of the perturbative regime and for very small $q$ the
anomalous scaling $G_{hh}\propto q^{-4+\eta}$ where $\eta\approx
0.85$ agrees with the NPRG result.

An important scale is the Ginzburg
scale $q_{G}$ for the crossover from the perturbative to the
anomalous regime. This scale cannot be captured within
a finite order derivative
expansion. Perturbation
theory \cite{membranebook,Nelson87} yields for $D=2$ the rough estimate 
\begin{equation}
q_G^{pt}\approx [3\tilde{K}_0/(2\pi)]^{1/2}/(4 {\tilde \kappa})
\label{eq:ginz}
\end{equation}
with $\tilde{K}_0=4J^2{\tilde \mu}({\tilde
\mu}+{\tilde \lambda}) /(2 {\tilde \mu}+{\tilde \lambda})$. Below we
extract $q_G$ from the numerical flow which gives
a more accurate estimate.

The RG equations require the initial form of 
$\tilde{\kappa}_{q}, \tilde{\mu}_{q},
\tilde{\lambda}_q$ and the mean-field order parameter $J_{\Lambda_0}$ which
are defined at the UV cutoff $\Lambda_0$. While we cannot rule out a
$q$ dependence of the initial coupling functions, for simplicity and
in accordance with Eqs.~(\ref{eq:bend}, \ref{eq:stretch}) we
choose $q$-independent constant values for the initial form of the
coupling functions. 
To fix the initial value of $\kappa$
we use a value previously reported
in the literature, $\kappa=1.1\,$eV.\cite{Fasolino07,Los09} 
The elastic properties were studied in
detail in Ref.~[\onlinecite{Zackarchenko09}] and values for
the bulk modulus $B=J^2(\mu+\lambda)$ and 
the shear modulus, in our notation $J^2\mu$, were extracted for moderate system sizes.
For scales smaller than the Ginzburg scale, all elastic constants are
strongly cutoff dependent and in particular in the IR limit
${\tilde\lambda}_{q=0}$ and ${\tilde\mu}_{q=0}$ vanish as $\Lambda^{4-D-2\eta}$ whereas 
${\tilde\kappa}_{q=0}$
diverges as $\Lambda^{-\eta}$. Values of the elastic
constants of free standing graphene must therefore be understood as
valid only for a given system size (or IR cutoff). The system size
used in Ref.~[\onlinecite{Zackarchenko09}] 
are for $T=300$~K at the border of the 
perturbative
regime and the reported elastic constants are not yet
strongly renormalized and close to those at smaller scales. 
We therefore use these results to
fix the initial values of $J^2 \mu\approx 9.95 \, {\rm eV}\, {\text \AA}^{-2} $ and 
$J^2 \lambda \approx 2.41 \, {\rm eV}\, {\text \AA}^{-2} $. 
The initial
value of the order parameter $J_{\Lambda_0}^2 \approx 2.5$ is chosen to
give
the best overall agreement with the MC data. 
These values place graphene well inside the flat phase,
in accordance with numerical simulations which show no sign
of crumpling even at high temperatures \cite{Fasolino07}.
The NPRG result for $G_{hh}$  with these 
values and $T=300$~K is
shown in Fig.~\ref{fig:Ghh}. 
The agreement with the MC data
is very good, especially the sharpness of the crossover from the
perturbative to the anomalous regime is well reproduced. 
A simple 
phenomenological ansatz for the self-energy,
$\Sigma_{hh}={\tilde \kappa}_{\Lambda_0} A \, q^4 (q_G/q)^\eta$ with $A$
fixed by the asymptotic behavior, does not lead to a satisfactory
description of the data, as
was already noted in Ref.~[\onlinecite{Los09}], see
Fig.~\ref{fig:Ghh}.
The Ginzburg scale is by standard definition the
scale where selfenergy correction to $\tilde{\kappa}_{q=0}$ equal
the bare parameter which allows to read $q_G\approx 0.08\, {\text \AA}^{-1}$
directly off the flow, which is slightly smaller than the perturbative
estimate $q_G^{pt}\approx 0.12\, {\text \AA}^{-1}$ 
from Eq.~(\ref{eq:ginz}).
The Ginzburg length 
$L_G=2\pi/q_G \approx
80\, {\text \AA}$ is of the same order as experimentally observed ripples 
\cite{Meyer07} which
offers a natural explanation of their appearance as just 
the real-space manifestation of the Ginzburg scale. Furthermore,
experimentally the fluctuations were found to be broadly distributed
around a characteristic scale, which is again in accordance with
the behavior of $G_h$ around $q_G$ which is not characterized
by a sharp feature at $q_G$ but by a gradual crossover from
the perturbative to the anomalous regime. The qualitatively
correct perturbative estimate of the Ginzburg scale
Eq.~(\ref{eq:ginz}) furthermore yields a simple dependence
of the Ginzburg scale on the temperature, $q_G \sim \sqrt{K_0 k_B T}/\kappa$.
If our interpretation of ripples in free standing graphene
is correct, the average real space scale of ripples should 
thus increase
as $\kappa/\sqrt{K_0 k_B T}$ on lowering the temperature, where
$K_0$ is the two dimensional Young's modulus.

\section{Conclusions}

In summary, we have presented a NPRG analyis for
tethered membranes which avoids a derivative expansion and
 is the first to include the full momentum
dependence of the elastic coupling parameters. Our solution of
the NPRG flow equations is completely self-consistent and obeys
all symmetry constraints. In our approach the crumpling transtion
is found to be of second order and
we give an improved estimate for the anomalous dimension at
the transition. 
For the flat phase of the membrane, 
we find excellent agreement with MC results for the momentum dependence of the
out-of-plane fluctuations
of free standing graphene. Also the crossover region, which
shows a relatively sharp crossover from the perturbative regime
to the anomalous scaling regime which is characterized
by a large anomalous dimension, is well reproduced.
This strongly supports the notion
that free standing graphene behaves just as a tethered membrane,
albeit a very stiff one. 
The most important scale in the analysis of the
momentum dependence of the membrane fluctuations is the Ginzburg scale
which we find to be of the same order as the experimentally determined
characteristic size of ripples. 
The observation
of ripples at this scale 
should thus be looked at as a confirmation of the continuum
elastic theory of tethered membranes, a notion which could
also be tested experimentally by measuring the characteristic
ripple scale as a funtion of temperature.

We thank Annalisa Fasolino, Jan H.~Los, Konstantin Zakharchenko, and
Misha Katsnelson for discussions and 
sharing their MC data with us prior
to publication. We further thank
Antonio H.~Castro Neto and Dominique Mouhanna for discussions.

\end{document}